\documentclass[preprintnumbers, floatfix, twocolumn,
preprintnumbers, PRD, superscriptaddress,nofootinbib]{revtex4-1}
\usepackage{eurosym}
\usepackage{amsfonts}
\usepackage{amsmath}
\usepackage{amssymb,epsf}
\usepackage{latexsym}
\usepackage{graphicx,epsfig}
\usepackage{amssymb}
\usepackage{subfigure}
\usepackage{epstopdf}
\usepackage[colorlinks=true,citecolor=blue,linkcolor=blue,urlcolor=black]{hyperref}
\usepackage{color}
\usepackage{float}
\usepackage{amsmath}
\graphicspath{{Images/}}

\renewcommand{\&}{\textup{\symbol{`\&}}}
\usepackage{hyperref}
\usepackage{amsfonts}
\usepackage{amsmath}

\usepackage{amssymb}
\usepackage{graphicx}%
\usepackage{bigints}
\usepackage{graphicx}
\usepackage{subfigure}
\usepackage{epsf}
\usepackage{bm}
\usepackage{amsmath}
\usepackage{amsfonts}
\usepackage{amssymb}
\usepackage{graphicx}
\usepackage{tabularx}
\usepackage{color}%
\providecommand{\U}[1]{\protect\rule{.1in}{.1in}}
\newcommand{\be}{\begin{equation}}
\newcommand{\ee}{\end{equation}}

\newcommand{\mincir}{\raise
-3.truept\hbox{\rlap{\hbox{$\sim$}}\raise4.truept\hbox{$<$}\ }}
\newcommand{\magcir}{\raise
-3.truept\hbox{\rlap{\hbox{$\sim$}}\raise4.truept\hbox{$>$}\ }}

\providecommand{\U}[1]{\protect\rule{.1in}{.1in}}

\usepackage{tikz,xcolor,hyperref}

\begin{document}

\title{Apparent dark matter as a non-local manifestation of emergent gravity}
\author{Kimet Jusufi}
\email{kimet.jusufi@unite.edu.mk}
\affiliation{Physics Department, State University of Tetovo, Ilinden Street nn,
1200,
Tetovo, North Macedonia}

\author{Ahmad Sheykhi}
\email{asheykhi@shirazu.ac.ir} \affiliation{Physics
Department and Biruni Observatory, College of Sciences,\\
Shiraz University, Shiraz 71454, Iran}

\author{Salvatore Capozziello}
\email{capozziello@unina.it}
\affiliation{Universit\`a degli studi di Napoli "Federico II", Dipartimento di Fisica "Ettore Pancini" and 
Istituto Nazionale di Fisica Nucleare, Sezione di Napoli, Complesso Universitario di Monte S.
Angelo, Via Cintia Edificio 6, 80126 Naples, Italy,\\
Scuola Superiore Meridionale, Largo S. Marcellino 10, 80138 Naples, Italy.
}

\begin{abstract}
We disclose a close correspondence between Verlinde's Emergent
Gravity (VEG) theory and the non-local gravity theories. Such
non-local effects can play crucial role at small distances as well
as in large scale structures. In particular, we argue that the
emergent gravity effectively is a manifestation of the
entanglement entropy and can modify Newton's law of gravity as
well as address the flat rotation curves of spiral galaxies.  In
the cosmological setup, we have considered three different models
for the apparent dark matter density. In the first model, we have
found that Friedmann equations get modified due to the presence of
the apparent dark matter (DM) in such a way that Newton's constant
of gravity shifts as $G\rightarrow G\left(1+\zeta\right)$, where
$\zeta$ is a dimensionless small parameter. Using the flat
rotating curves we estimate $\zeta \sim 10^{-7}$. For such a
model, we find out that by rescaling the radial coordinate, $r$,
the curvature space constant, $k$, and the scale factor of the
universe, $a$, the effect of apparent DM can change the geometry
of the universe and can shift the curvature space constant as
$k_{\star}=k (1-\zeta)$. To this end, we study a
more realistic model with evolving densities where we obtain the
$\Lambda$CDM model in Verlinde's gravity and comment on the Hubble
tension problem with evolving densities.
\end{abstract}
\maketitle

%%%%%%%%%%%%%%%%%%%%%%%%%%%%%%%%%%%%%%%%%%%%%%%%%%%%%%%%%%%%%%%%%%%%%%%%%%%
\section{Introduction}
Thermodynamics of black holes implies that there should be a deep
connection between gravity and thermodynamics. In 1995, Jacobson
\cite{Jac} claimed that the Einstein field equations of general
relativity is nothing but an equation of state for the spacetime.
Starting from fundamental relation $\delta Q=T \delta S$, together
with the area law of entropy, $S\sim A/4$, he derived the
hyperbolic second order partial differential equations of general
relativity in its covariant form. Soon after Jacobson, many
investigations have confirmed that there is indeed a deeper
connection between gravitational field equations and the laws of
thermodynamics \cite{Elin,Pad,CaiKim,Cai1,Cai2,
Wang,Shey1,Shey2,Shey3,Shey4}. The deep connection between the
gravitational equation describing the gravity in the bulk and the
first law of thermodynamics on the boundary of spacetime reflects
some deep ideas of holography. Although Jacobson's derivation is
logically clear and theoretically sound, the statistical
mechanical origin of the thermodynamic nature of general
relativity remains obscure.

The next step towards understanding the nature and origin of
gravity put forward by Verlinde \cite{Verlinde} who suggested that
gravity is not a fundamental force and can be regarded as an
emergent phenomena. He claimed that the laws of gravity emerge as
an entropic force caused by the changes in the information
associated with the positions of material bodies. According to
Verlinde argument when a test particle moves apart from the
holographic screen, the magnitude of the entropic force on this
body has the form
\begin{equation}\label{F}
F\triangle x=T \triangle S,
\end{equation}
where $\triangle x$ is the displacement of the particle from the
holographic screen, while $T$ and $\triangle S$ are, respectively,
the temperature and the entropy change of the screen.
According to the holographic description, to
recover the gravity we  have to assume that information is stored
on surfaces, or screens. Using the entropic force scenario as
well as the equipartition law of energy and the holographic
principle, Verlinde derived Newton's law of gravitation, the
Poisson equation and in the relativistic regime, the Einstein
field equations of general relativity \cite{Verlinde}. The
Verlinde's viewpoint on the nature of gravity opened a new window
to understand gravity from the first principles. The study on the
entropic force has raised a lot of attention in the literature
(see \cite{Cai3,Other,newref,sheyECFE,Ling,Modesto,Yi,Shey5} and
references therein).

On the other side, the DM puzzle has been one the most important
challenges of the modern astrophysics. The problem originates from
the fact that observations confirm that the rotational velocity
curves of all spiral galaxies tend to some constant value
\cite{Rubin,Trimble,Bahcall,Vogt}. These observations are in
contradiction with the prediction of Newtonian theory because
Newtonian theory predicts that objects that are far from the
galaxy center have lower velocities. The most widely adopted way
to resolve these difficulties is the DM hypothesis. It is assumed
that all visible stars are surrounded by massive nonluminous
matters. Another approach is the Modified Newtonian dynamics
(MOND) theory which was proposed by Milgrom in 1983
\cite{Milgrom}. The MOND theory has been successful for explaining
the observed anomalous rotational-velocity. Although MOND theory
can explain the flat rotational curve, however its theoretical
origin remains un-known. Inspired by the Debye model for the
equipartition law of energy in statistical thermodynamics and
adopting the viewpoint that gravitational systems can be regarded
as a thermodynamical system, it was argued that the Debye entropic
gravity can explain the theoretical origin of the MOND theory
\cite{SheyMOND}. Many authors have tried to explain the flat
rotation curves through modification of Einstein gravity
\cite{Sobuti1,Sobuti2,Mim1,Myr2,MimMOND,ASJ,SheyTs}.

Following the theoretical progress which indicates that spacetime
and gravity can be emerged, in another paper
\cite{Verlinde:2016toy} Verlinde argued that the volume law
contribution to the entanglement entropy, associated with the
positive dark energy, turns the otherwise ``stiff'' geometry of
spacetime into an elastic medium. It was shown that the elastic
response of this ``dark energy'' medium takes the form of an extra
``dark'' gravitational force called apparent DM
\cite{Verlinde:2016toy}. The black hole solutions in this theory has been studied in \cite{jusufi}. In this paper, by taking into account the
zero-point length correction to the gravitational potential, we
modify the apparent DM hypothesis as well as Newton's law of
gravity.  We also address the correspondence between Verlinde's
gravity and the nonlocal gravity. Finally, we extend our study to
the cosmological setup. Through this paper we shall set
$c=\hbar=1$.

This paper is structured as follows.  In the next section we
present an expression for apparent DM in the context of VEG
theory. In section III,  we point out the correspondence to non-local gravity. In section IV, we investigate the consequences of the
apparent DM on the cosmological field equations. We shall then
obtain the $\Lambda$CDM model in this
scenario in section IV. In section V, we study  a general model with evolving densities. We finish our paper with conclusions which
appear in section VI.
%%%%%%%%%%%%%%%%%%%%%%%%%%%%%%%%%%%%%%%%%%%%%%%%%%%%%%%%%%%%%%%%%%%%%%%%%%%%%%%%%%%%%%%
\section{Modified apparent DM from emergent gravity}
A wide range of observational data indicate that the rotational
velocity curves of all spiral galaxies are proportional to the
distance from the center $v\propto r$, for inside the galaxy, and
remain \textit{almost} flat at large distance from the galaxy
center, namely at the galaxy out skirt, typically beyond $30-40$
kpc \cite{Rubin,Trimble,Bahcall,Vogt}. While inside the galaxy,
these observations consist with Newtonian gravity, however, for
outside a galaxy, Newton's law fails to explain the rotational
curves. In fact, Newton's law of gravitation predicts that objects
that are far from the galaxy center have lower velocities $v
\propto 1/\sqrt{r}$, in contradiction with observations which
indicate, $v\simeq \rm constant$ for outside of galaxy. It is a
general belief that the baryonic matter of galaxies does not
provide sufficient gravitation to explain the observed dynamics of
the systems. In this section, by modifying Newton's law of
gravity, we address the puzzle of flat rotation curves via
apparent DM hypothesis through emergent gravity theory.

According to Verlinde's argument, by assuming a spherical
symmetry, the amount of apparent DM, $M_D(r)$, is related to the
amount of baryonic matter $M_B(r)$ ~\cite{Verlinde:2016toy}. To
see this, we can write the integral relation for the surface mass
density $\sigma_D$ of the apparent DM in terms of the Newtonian
potential for the baryonic matter as \cite{Verlinde:2016toy}
\begin{eqnarray} \label{sigma}
\int_{\mathcal{B}} \left(\frac{8 \pi G }{a_0}
\sigma_D\right)^2dV=\frac{d-2}{d-1}\oint_{\partial
\mathcal{B}}\frac{\Phi_B}{a_0}n_i dA_i,
\end{eqnarray}
where $a_0$ is the acceleration scale and plays an important role
in Verlinde's emergent theory and is related to the Hubble
parameter with $a_0=c H_0$. Furthermore, $\Phi_B$ is the Newtonian
potential due to the baryonic matter, $n_i$ and $dA_i$ are the
unit vector and area element yielding $dA=n_i dA_i$. For a
spherically symmetric situation this relation, the surface mass
density is defined as
\begin{eqnarray}
    \sigma_D(r)=\frac{M_D(r)}{A(r)}.
\end{eqnarray}
In addition, we use the regularized expression for the gravitational potential
\cite{td1,td2,Jusufi:2022mir}
\begin{eqnarray} \label{PhiB}
    \Phi_B=-\frac{G M }{\sqrt{r^2+l_0^2}},
\end{eqnarray}
where $M$ is the bare mass of the baryonic mass in the system and
$l_0$ is the zero-point length and its value is of order of Planck
length \cite{td1}. Inserting relation (\ref{PhiB}) into Eq.
(\ref{sigma}), for $d=4$, we find
\begin{equation}\label{IntM}
 \int_0^r\frac{G M^2_D(r^\prime)}{{r^\prime}^2}dr^\prime = \frac{cH_0 M}{6}\frac{r^2}{\sqrt{r^2+l_0^2}},
\end{equation}
which reduces to the Verlinde's formula \cite{Verlinde:2016toy} in
the limit $l_0\to 0$, namely
\begin{equation}
 \int_0^r\frac{G M^2_D(r^\prime)}{{r^\prime}^2}dr^\prime = r\frac{cH_0M}{6},
\label{Verlinde's formula}
\end{equation}
provided we have a constant point-like mass $M=M_B$ and
$H_0$ is the Hubble parameter, $\Lambda$ is the cosmological
constant $G=6.674\times 10^{-11}~{\rm m^3/(kg s^2)}$ is the
Newton's constant, $c\approx 3\times 10^8~{\rm m/s}$ is the speed
of light. We can also rewrite Eq. (\ref{IntM}) as follows
\begin{equation}
\int_0^r \frac{G M^2_D(r') dr'}{r'^2}=\frac{a_0
M}{6}\frac{r^2}{\sqrt{r^2+l_0^2}}.
\end{equation}
Furthermore, one can express this relation as
\begin{equation}
M_D^2(r)=\frac{a_0
r^2}{6\,G}\frac{d}{dr}\left[\frac{r^2}{\sqrt{r^2+l_0^2}} M\right].
\end{equation}
Hereinafter we will also use \cite{Verlinde:2016toy}
\begin{equation}
a_M=\frac{a_0}{6},
\end{equation}
where $a_0=5.4 \times 10^{-10} \text{m/s}^2$ comes from
observations (see also \cite{Milgrom}). Since $M$ is a constant,
we can introduce $\xi=\sqrt{a_M M G}$, The most natural regular
black hole solution with stringy effects can be found by using the
potential (3) and solve the Poisson's equation
\begin{equation}
\nabla^2 \Phi_B(r)=4 \pi G \rho_B(r).
\end{equation}
This leads to the following mass function
\begin{eqnarray}
M_B(r)=4\pi \int_0^r r'^2 \rho_B(r') dr'=\frac{
M\,r^3}{(r^2+l_0^2)^{3/2}}.
\end{eqnarray}
Using this equation, we can also express the relation between
apparent DM mass and the baryonic mass as follows
\begin{equation}
M_D^2(r)=\frac{a_M}{G} r^2\frac{d}{dr}\left[\frac{r^2+l_0^2}{r} M_B(r)\right].
\end{equation}
When $l_0\to 0$, we obtain the Verlinde's formula for the apparent
DM. Using Eq. (10), we obtain the corresponding apparent DM mass
as
\begin{equation}
%\xi r \sqrt{\frac{r (2
%l_0^2+r^2)}{(r^2+l_0^2)^{3/2}}}
M_D(r)\simeq \frac{\xi\, r}{G}
\left(1+\frac{l_0^2}{4\,r^2}\right)+....
\end{equation}
where  we have kept only terms up to order $l_0^2/r^2$. Using the
mass of the apparent DM, we can get the apparent DM density
\begin{equation}
 \rho_D(r)=\frac{1}{4 \pi r^2}\frac{dM_D(r)}{dr}\simeq \frac{\xi}{4 \pi G r^2}+....
\end{equation}
For astrophysical purposes, only the first term is important as $l_0$ is of the Planck order length. Substituting $\rho_D(r)$ in the Poisson's equation,
\begin{equation}
\nabla^2 \Phi_D(r)=4 \pi G \rho_D(r),
\end{equation}
and solving the differential equation for the radial part of the Laplacian we obtain the gravitational potential
\begin{equation}
 \Phi_D= \xi  \ln (r)-\frac{ \xi \,l_0^2}{8 r^2}-\frac{c_1}{r}+c_2.
\end{equation}
where $c_1$ and $c_2$ are integration constants. Since as
$\xi\rightarrow 0$, we require to have $\Phi_D=0=M_D$, thus
$c_1=c_2=0$.

%\textcolor{red}{On the other hand if we take $c_2=0$ and
%$c_1=M_B$, then $\Phi_D$ contain both DM and BM terms in it and
%one do not need to introduce $\Phi_B$, separately, what do you
%think?? }

The total force acting on a test particle with mass $m$ is
therefore given by
\begin{equation}
 F= F_B+ F_D,
\end{equation}
here $F_D$ is the force acting on the particle due to the dark
companion. In particular we have
\begin{equation}
F_B=-m\nabla \Phi_B(r)=-\frac{G M m\, r}{(r^2+l_0^2)^{3/2}},
\end{equation}
and
\begin{equation}
F_D=-m\nabla \Phi_D(r)\simeq - \frac{ \xi
m}{r}\left(1+\frac{l_0^2}{4 r^2}\right).
\end{equation}

%\textbf{From this result and using dimensional arguments, I expect
%$\xi\sim M_D$, while already you defined $\xi=\sqrt{a_M M}$! which
%means the unit of $\xi$ is $\sqrt{F}$. Something is wrong...PLs
%check the units.. }

For the total force we obtain
\begin{equation}
F=- \frac{ G M m r}{(r^2+l_0^2)^{3/2}}- \frac{ \xi
m}{r}\left(1+\frac{l_0^2}{4 r^2}\right).
\end{equation}
Since $l_0$ is very small, we can approximate the above force in
leading order terms
\begin{equation}\label{FM}
F\simeq - \left[\frac{ G M m }{r^2}\left(1-\frac{3 l_0^2}{2
r^2}\right)+\frac{\xi m}{r}\left(1+\frac{l_0^2}{4 r^2}
\right)\right].
\end{equation}
Therefore, the Newtonian dynamics get modified due to the
zero-point length corrections as well as apparent DM. One can
easily check that the modified Newtonian dynamics obtained in Eq.
(\ref{FM}) can explain the flat rotation curves of galaxies. Using
the fact that $|F|=m v^2/r$, we can rewrite the circular speed of
an orbiting test object as
\begin{equation}
v^2 \simeq \left[\frac{ G M }{r}\left(1-\frac{3 l_0^2}{2
r^2}\right)+ \xi \left(1+\frac{l_0^2}{4 r^2} \right)\right].
\end{equation}
As a result, for large values of $r$, namely out of galaxies, $v^2
\approx \xi=\sqrt{a_M M G}=constant$, which implies that the
rotation curves of galaxies are flat, which confirm by
observations \cite{Rubin}. It is important to note that zero-point
length corrections to the Newton's law of gravity, do not play
significant role at large distance, while apparent DM term,
$\xi=\sqrt{a_M M G}$, plays crucial role at large distances.
%%%%%%%%%%%%%%%%%%%%%%%%%%%%%%%%%%%%%%%%%%%%%%%%%%%%%%%%%%%%%%%%%%%%%%%%%%%%%%%%%%%%%%%%%%%%%%%%
\section{Correspondence between Verlinde's gravity and nonlocal gravity}
%\textcolor{blue}{This section needs to be written and imporved}
For large distance ($l_0\ll r$), we can only consider the
logarithmic correction in Eq. (16) to the gravitational potential
and write
\begin{equation}
    \Phi_D\simeq\xi \ln r.
\end{equation}
On the other hand, for the gravitational potential due to the
baryonic matter from Eq. (4), we get
\begin{equation}
    \Phi_B \simeq -\frac{G M }{r}.
\end{equation}
Again, we have neglected the quantum correction term due to $l_0$.
For the total gravitational potential, we get
\begin{equation}\label{Phi}
    \Phi\simeq -\frac{G M }{r}+\xi \, \ln \frac{r}{\mathcal{C}}.
\end{equation}
where $\mathcal{C}$ is introduced to obtain a dimensionless quantity for $r/\mathcal{C}$.
It is interesting to point out that such logarithmic corrections
are characteristic a non-local theory of gravities
\cite{ng1,ng2,ng3}.  One might ask now if the correspondence
between Verlinde's emergent gravity and non-local gravity signals
a deep connection between the nature of gravity and nonlocality.
To quote Verlinde  \cite{Verlinde:2016toy} \textit{''...DM is
consequence of the emergent nature of gravity and caused by an
elastic response due to the volume law contribution to the
entanglement entropy in our universe''}. If apparent DM is
described by entanglement entropy in our universe as Verlinde
says, we know that entanglement is a non-local behavior of quantum
mechanics, then there is no surprise to say Verlinde's theory is a
non-local gravity theory too.

We will argue that there are two typos of non-localities in the present paper.
\begin{itemize}
\item  Short-range nonlocality
\end{itemize}
It has to do with the fact that the point-like nature of particles
is not accurate and we need to replace it with a smeared matter
distribution. The short range non-locality is encoded in the
zero-point length parameter $l_0$ which has the length of Planck
length order. One way to see this is rewrite Eq. (11) by restoring
the Newton constant
\begin{eqnarray}
M_B(r)=G(r) M,
\end{eqnarray}
where $G(r)$ incorporates all the effects of non-locality and it is given by (see \cite{td1})
\begin{eqnarray}
G(r)=\frac{r^3}{(r^2+l_0^2)^{3/2}},
\end{eqnarray}
which is a feature of non-locality due to the smeared matter distribution. The expression is similar to the expression
for the running of the gravitational constant in asymptotically safe gravity \cite{Bonanno:2000ep}.

\begin{itemize}
\item  Long-range non-locality
\end{itemize}
As we pointed out, according to the nonlocal gravity theory, non-locality persists in the
Newtonian regime as well. Indeed, the Poisson's equation of Newtonian theory
of gravitation is linearly modified as was argued in Refs. \cite{ng1,ng2,ng3}.
In this case, the nonlocal gravitational field equations reduce to
\cite{ng1,ng2,ng3}
\begin{equation}\label{Ba2}
  G_{\mu\nu}(x)+\int{\cal K}(x,y)G_{\mu\nu}(y)d^4y=\kappa T_{\mu\nu}\,,
\end{equation}
in which $G_{\mu\nu}$ is the linear
Einstein tensor and $T_{\mu\nu}$ is the
energy-momentum tensor of matter. One can fursther introduce the reciprocal
kernel ${\cal R}(x,y)$ via
\begin{equation}\label{Ba4}
-{\cal R}(x,y)=\sum_{n=1}^\infty {\cal K}_n (x,y)\,.
\end{equation}
and rewrite Eq.\ (\ref{Ba2})  as follows
\begin{equation}\label{Ba5}
G_{\mu\nu}(x)= \kappa\left[ T_{\mu\nu}(x) +\int{\cal R}(x,y)
T_{\mu\nu}(y)d^4y\right]\,,
\end{equation}
and we end up with an interesting consequence that that the nonlocal theory in this approximation is equivalent to GR
but with an additional source term. It was argued that each iterated kernel is proportional to $\delta(x^0-y^0)$ and in particular  \cite{ng1,ng2,ng3}
\begin{eqnarray}
{\cal R}(x,y)=\delta(x^0-y^0)q(\mathbf{x}-\mathbf{y})\,,
\end{eqnarray}
where $q$ is the spatial convolution kernel. Using this limiting form
of the kernel in Eq.\ (\ref{Ba5}), we find in the Newtonian limit the
nonlocal Poisson equation  \cite{ng1,ng2,ng3}
\begin{equation}\label{Newton}
\nabla^2\Phi=4\pi G\left[\rho(t,\mathbf{x})+\rho_{\rm D}(t,\mathbf{x})\right]\,,
\end{equation}
where the ``density of DM'' $\rho_{\rm D}$ is given by
\begin{equation}\label{dark}
\rho_{\rm D}(t,\mathbf{x})=\int q(\mathbf{x}-\mathbf{y})
\rho(t,\mathbf{y})d^3y\,.
\end{equation}
In the last equation $q$ is a universal function that is
independent of the nature of the source. It was argued that the
nonlocal aspect of the gravitational interaction acts like DM of
density $\rho_{\rm D}$ that is linearly related to the actual
matter density via the kernel $q$ as in equation.  For the case of
circular motion of stars we can explain the observed flat rotation
curves in galaxies using the kernel \cite{ng1,ng2,ng3}
\begin{equation}\label{qernel}
q(\mathbf{x}-\mathbf{y})=\frac{1 }{4\pi \lambda}
\frac{1}{|\mathbf{x}-\mathbf{y}|^2}\,,
\end{equation}
where $\lambda=GM/v^2$. The universality
of the nonlocal kernel implies that $\lambda$ must be a constant and
hence $M\propto v^2$. In particular, the force of gravity per unit
test mass is given by
\begin{equation}
\Phi=-\frac{G M }{r}+ \frac{GM}{\lambda}\, \ln(\frac{r}{\lambda}).
\end{equation}
which coincides with the results obtained from
Verlinde's emergent gravity for $\xi=GM\lambda^{-1}$. Now one is
capable to explain the flat rotation curves at large distances,
namely out of galaxies. For the circular speed of an orbiting test
object is given by (see also \cite{Sobuti1})
\begin{equation}\label{v2}
   v^2=\frac{G M}{r}+\xi.
\end{equation}
By Tully-Fisher relation, the asymptotic speed of a test particle,
becomes $v\sim \sqrt{\xi}$, which is consistent with
observations.  We can therefore say that non-local gravity can act
both at UV and IR scales and this fact is extremely relevant  in order to
fix the issue of a relation between general relativity  and quantum mechanics.
The main point here might be that as soon as we  "localize" the non-linear gravity, corrections immediately arise and we have the effect of a scalar field ruling the gravitational interaction. This
scalar tensor picture can rule gravitational interaction at various scales
ranging from clusters of galaxies \cite{Salv1,Salv2} and cosmology \cite{Salv3,Salv4}. Let us mention here that cosmology in non-local gravity has been studied also in \cite{Calcagni:2013vra}. Besides, the large
scale structure with logarithmic effects in non-linear gravity theory are studied in
\cite{Capozziello}.  These effects they are related to the existence of some Noether symmetry as shown in \cite{Salv5} and, as we pointed out strictly related to the amount of dark matter \cite{Salv6}. This could mean that the that emergent gravity effectively can be viewed as non-linear gravity and they are strictly related in some sense about the prediction of dark matter. Like so, the MOND could be related to $f(R)$. and non-linear gravity can be always represented as $f(R,\phi)$ which supports the above arguments.

%%%%%%%%%%%%%%%%%%%%%%%%%%%%%%%%%%%%%%%%%%%%%%%%%%%%%%%%%%%%%%%%%%%%%%%%%%%%%%%%%%%%%%%%%%%%%%
\section{Modified Friedmann equations}
Let us now extend our discussion to the cosmological setup.
Assuming the background spacetime is spatially homogeneous and
isotropic which is described by the FRW metric,
\begin{equation}
ds^2={h}_{\mu \nu}dx^{\mu} dx^{\nu}+R^2(d\theta^2+\sin^2\theta
d\phi^2),
\end{equation}
where $R=a(t)r$, $x^0=t, x^1=r$, and the two dimensional metric is
\begin{equation}
h_{\mu \nu}=diag(-1, a^2/(1-k r^2)).
\end{equation}
In this equation $k$ denotes the curvature of space with $k = 0, 1, -1$
corresponding to flat, closed, and open universes, respectively.
The dynamical apparent horizon, a marginally trapped surface with
vanishing expansion, is determined by the relation
\begin{eqnarray}
h^{\mu \nu}\partial_{\mu}R\partial_{\nu}R=0.
\end{eqnarray}
A simple calculation gives the apparent horizon radius for the FRW
universe
\begin{equation}
\label{radius}
 R=ar=\frac{1}{\sqrt{H^2+k/a^2}}.
\end{equation}
For the matter source in the FRW universe we shall assume a perfect
fluid described by the stress-energy tensor
\begin{equation}\label{T}
T_{\mu\nu}=(\bar{\rho}+p)u_{\mu}u_{\nu}+pg_{\mu\nu},
\end{equation}
where $u_{\mu}$ is the 4-velocity vector of the
perfect fluid, and $\bar{\rho}=\rho_B+\rho_D+\rho_\Lambda$ is the
sum of baryonic matter, apparent DM and cosmological constant
energy density, respectively. Having in mind the energy-momentum
conservation of matter $\nabla_{\mu}T^{\mu\nu}=0$, we obtain the
continuity equation
\begin{equation}\label{Cont}
\dot{\bar{\rho}}+3H(\bar{\rho}+p)=0,
\end{equation}
where $H=\dot{a}/a$ being the Hubble parameter. Let us now derive
the dynamical equation for Newtonian cosmology. Toward this goal,
let us consider a compact spatial region $V$ with a compact
boundary $\mathcal S$, which is a sphere having radius $R= a(t)r$,
where $r$ is a dimensionless quantity. Going back and combining
the second law of Newton for the test particle $m$ near the
surface, with gravitational force  we obtain
\begin{eqnarray}\label{F6}
F=m\ddot{R}&=&-\Bigg{\{}\frac{G M_B  m }{R^2}+\frac{\xi  m}{R}
\frac{\left(1+\frac{l_0^2}{4R^2}\right)}
{\left(1+\frac{l_0^2}{R^2}\right)^{-3/2}}\Bigg{\}}\nonumber
\\&& \times \left(1+\frac{l_0^2}{R^2}\right)^{-3/2}.
\end{eqnarray}
In what follows, we shall neglect the second order correction
terms, $\xi l_0^2$, hence from Eq. (\ref{F6}) we get
\begin{equation}\label{F7}
\ddot{a} r=-G\left[\frac{M_B+M_D}{a^2 r^2}\right]\left(1+\frac{l_0^2}{a^2 r^2}\right)^{-3/2},
\end{equation}
where the apparent DM mass is given by
\begin{equation}
M_D=\frac{\xi a r}{G}.
\end{equation}
In order to derive the Friedmann equations of FRW universe in
general relativity, we can use the active gravitational mass
$\mathcal M$, rather than the total mass $M$.
\begin{equation}
\mathcal M=M_B+M_D=M_B+\frac{\xi a r}{G}.
\end{equation}
It follows that, due to the
entropic corrections terms via the zero-point length, the active gravitational mass
$\mathcal M$ will be modified. Using Eq.
(\ref{F7}) and replacing $M$ with $\mathcal M$, it follows
\begin{equation}\label{M1}
\mathcal M =-\frac{\ddot{a}
a^2}{G}r^3\left(1+\frac{l_0^2}{R^2}\right)^{3/2}.
\end{equation}
In addition, for the active gravitational mass we can use the definition
\begin{equation}\label{Int}
\mathcal M =2
\int_V{dV\left(T_{\mu\nu}-\frac{1}{2}Tg_{\mu\nu}\right)u^{\mu}u^{\nu}}.
\end{equation}
Substituting the stress-energy tensor (\ref{T}) in
Eq. (\ref{Int}), and using the fact that $T=g^{\mu \nu}T_{\mu
\nu}$ and $u_{\mu}u^{\mu}=-1$, one can arrive at
\begin{equation}\label{M2}
\mathcal M =(\bar{\rho}+3p)\frac{4\pi}{3}a^3 r^3.
\end{equation}
Equating Eqs. (\ref{M1}) and (\ref{M2}), one gets
\begin{equation}\label{addot}
\frac{\ddot{a}}{a} =-\frac{4\pi  G
}{3}(\bar{\rho}+3p)\left[1+\frac{l_0^2}{R^2}\right]^{-3/2}.
\end{equation}
This is the modified acceleration equation for the dynamical
evolution of the  FRW universe.
For the Friedmann equation we therefore obtain
\begin{equation}\label{addot0}
\frac{\ddot{a}}{a}=- \left(\frac{4 \pi G }{3}\right)
(\bar{\rho}+3p)\left[1-\frac{3}{2} \frac{l_0^2}{r^2 a^2}+... \right] .
\end{equation}
Next, by multiplying $2\dot{a}a$ on both sides of Eq.
(\ref{addot}), and by means of continuity equation (\ref{Cont}),
we obtain
\begin{equation}\label{Fried1}
\dot{a}^2+k =\frac{8\pi G
}{3}\int d(\bar{\rho} a^2)\left[1-\frac{3}{2} \frac{l_0^2}{r^2 a^2}\right],
\end{equation}
where $k$ is a constant of integration and physically characterizes the curvature of space. Using the expression for density
\begin{eqnarray}
\bar{\rho}\equiv \sum_i \rho_{0i} a^{-3 (1+\omega_i)},
\end{eqnarray}
which, in the general case, includes several matter fluids. We obtain 
\begin{equation}\label{Fried2}
H^2+\frac{k}{a^2} =\frac{8\pi G }{3} \sum_i \rho_{i}\left[1-\sum_i\Gamma(\omega_i) \rho_i \right],
\end{equation}
where $\Gamma$ is a constant defined as
\begin{equation}\label{Gamma}
\Gamma (\omega_i)\equiv\frac{4\pi G l_0^2 }{3}\left(\frac{1+3
\omega_i}{1+\omega_i}\right).
\end{equation}

 Before we consider in more details how apparent dark matter effects the Friedmann equation, let us introduce the average mass densities $\rho_B(r)$ and $\rho_D(r)$
inside a sphere of radius $R$ by writing the integrated masses. The total mass density profile inside the core is
then given by
\begin{eqnarray}
M_B(R)&=&\frac{4 \pi}{3}R^3 \rho_B(R),\\
M_D(R)&=&\frac{4 \pi}{3}R^3 \rho_D(R)
\end{eqnarray}

Following Verlinde (\cite{Verlinde:2016toy}), we can introduce the slope parameters
\begin{eqnarray}
\bar{\beta}_B(R)&=&-\frac{d\log \rho_B(R)}{d \log R}, \\
\bar{\beta}_D(R)&=&-\frac{d\log \rho_D(R)}{d \log R}.
\end{eqnarray}
respectively. On the other hand, from the relation between the apparent dark matter and baryonic matter (12) one can find the relation between the densities
\begin{equation}
\rho^2_D(R)=(4-\bar{\beta}_B(R))\frac{a_0}{8 \pi G R}\rho_B(R)
\end{equation}
where we assumed $l_0 \to 0$ for cosmological scales.
%Finally, given the average mass density $\bar{\rho}_D(r)$ one can find the actual mass density $\rho_D(r)$
%for apparent dark matter via the relation
%\begin{equation}
%\rho_D(r)=(1-\frac{1}{3}\bar{\beta}_D(r))\bar{\rho}_D(r).
%\end{equation}

\subsection{Case $\bar{\beta}_B=3$ and $ \bar{\beta}_D=2$}

Let us now focus on a particular example by taking:  $\bar{\beta}_B=3, \bar{\beta}_D=2$. In this way, it follows that
\begin{equation}
\rho_D=\frac{C}{R^2},
\end{equation}
where $C$ is a constant. Taking the constant $C=\xi/4 \pi G$, we obtain
\begin{equation}
\rho_D=\frac{\xi}{4 \pi G R^2},
\end{equation}
which is Eq.(14) in the present paper. For the Fredmann equation then we have
\begin{equation}\label{Fried2}
H^2+\frac{k}{a^2} =\frac{8\pi G }{3}\left[\rho_B+\rho_\Lambda \right]\left[1-\sum_i\Gamma(\omega_i) \rho_i \right]+\frac{8\pi G }{3}\rho_D,
\end{equation}
where $\rho_\Lambda=\Lambda/8\pi G$. If we now use $R=r a$ given by Eq. (40) we get
\begin{equation}\label{Fried2}
(H^2+\frac{k}{a^2})(1-\zeta) =\frac{8\pi G }{3}\left[\rho_B+\rho_\Lambda \right]\left[1-\sum_i\Gamma(\omega_i) \rho_i \right],
\end{equation}
where $\zeta=2 \xi/(3c^2)$ [here we have restored also speed of
light  constant $c$]. Note that $\zeta$ has no dimensions and we
expect $\zeta$ to be small number. In fact we can constrain
$\zeta$ using the flat rotating curved in galaxies $v^2\sim \xi$,
taking for example  a typical velocity of a star in galaxy to be
$v \sim (200-300)$km/s, we find
\begin{eqnarray}
\zeta \sim \frac{2 v^2}{3 c^2} \sim 10^{-7}.
\end{eqnarray}
In that case, we can rewrite equation (64) as follows
\begin{equation}\label{Fried2}
H^2+\frac{k}{a^2}=\frac{8\pi G (1+\zeta) }{3} \left[\rho_B+\rho_\Lambda \right]\left[1-\sum_i\Gamma(\omega_i) \rho_i \right].
\end{equation}
It is interesting
to observe that the effect of apparent DM in cosmological scales
shifts the Newton constant
\begin{equation}
 G_N=G\,(1+\zeta).
\end{equation}
However such a shift of Newton constant is found in MOG gravity theory \cite{Moffat:2020ffz}. Hence we can write the Friedmans equations
\begin{equation}\label{Fried2}
H^2+\frac{k}{a^2}=\frac{8\pi G_N }{3} \left[\rho_B+\rho_\Lambda \right]\left[1-\sum_i\Gamma(\omega_i) \rho_i \right].
\end{equation}

This is the corrected Friedman and the non-locality is encoded in the Newton
constant; the apparent DM dominates in large scale structures by
increasing the standard Newton's constant $G \to G(1+\zeta)$.

\subsubsection{An alternative explanation: Rescaling of coordinates}
There is, however, an alternative way to look at the effect of DM
in cosmology. To see this let us look more closely in the LHS of
the Friedman equation  (64) which reads
\begin{equation}\label{Fried2}
\left(H^2+\frac{k}{a^2}\right)\left(1-\zeta\right)=\frac{(1-\zeta)}{a^2 r^2},
\end{equation}
now if we introduce the following definitions
\begin{equation}
 r_{\star}=\frac{r}{\sqrt{1-\zeta}},
\end{equation}
\begin{equation}
 k_{\star}=k (1-\zeta),
\end{equation}
\begin{equation}
 a_{\star}(t)=a(t) \sqrt{1-\zeta},
\end{equation}
then, quite remarkably, this quantity remains invariant
\begin{equation}\label{Fried2}
H_{\star}^2+\frac{k_{\star}}{a_{\star}^2}=\frac{1}{a_{\star}^2 r_{\star}^2},
\end{equation}
and so the FRW metric
\begin{equation}
ds^2=-dt^2+a_{\star}(t)^2 \left[ \frac{dr_{\star}^2}{1-k_{\star}
r_\star^2} +r_{\star}^2 (d\theta^2+\sin^2\theta d\phi^2)  \right].
\end{equation}
written in these new coordinates. The metric of the FRW universe
is the same  as in general relativity, however the curvature space
constant is now shifted to $k_{\star}$. We shall elaborate the
implications of this later on. This means that we can write
\begin{equation}\label{Fried2}
H_{\star}^2+\frac{k_{\star}}{a_{\star}^2}=H^2+\frac{k}{a^2},
\end{equation}
and rewrite the Fredmann equation only in terms of baryonic matter and cosmological constant
\begin{equation}\label{Fried2}
H^2+\frac{k}{a^2}=\frac{8\pi G  }{3}\rho_B\left[1-\sum_i\Gamma(\omega_i) \rho_i\right]+\frac{\Lambda}{3}.
\end{equation}
For the case of vanishing $\Lambda$, this equation was recently derived in \cite{Jusufi:2022mir}. This indicates that the effect of apparent DM can be viewed either as a shift of the Newton constant provided the spatial curvature
constant is $k$, or, equivalently, we shift the curvature constant
$k_{\star}=k(1-\zeta)$ provided the Newton constant is unchanged.
Let us now see more closely the effect of apparent DM on the
geometry of the universe. First, if we fix the radial coordinate
$r=\rm const$, and constant time $t=\rm const$, we get
\begin{equation}
ds^2=a_{\star}^2r_{\star}^2 (d\theta^2+\sin^2\theta d\phi^2)
\end{equation}
where
\begin{eqnarray}
a_{\star}^2r_{\star}^2=a(t)^2 (1-\zeta)
\left[\frac{r^2}{(1-\zeta}\right]=a(t)^2 r^2.
\end{eqnarray}
Now we can use the Gaussian curvature with respect to $g^{(2)}$ on
$\mathcal{M}$ and the Gauss-Bonnet theorem to show that the
topology does not change. The Gauss-Bonnet theorem states that
\begin{equation}
\iint_{\mathcal{M}} K dA=2 \pi \chi(\mathcal{M}),
\end{equation}
where $dA$ is the surface line element of the 2-dimensional
surface and $\chi(\mathcal{M})$ is the Euler characteristic
number. It is convenient to express sometimes the above theorem in
terms of the Ricci scalar, in particular for the 2-dimensional
surface at $R|_{t=const, r=const}=a r=const$, then using the Ricci
scalar given by
\begin{equation}
\mathcal{R}=\frac{2}{R^2},
\end{equation}
with $ \sqrt{g^{(2)}}=R^2\sin\theta$, evaluated at $R$, yielding the following from
\begin{equation}
\frac{1}{4 \pi}\iint_{\mathcal{M}} \mathcal{R} \sqrt{g^{(2)}} \,d\theta d\phi=\chi(\mathcal{M}).
\end{equation}
This shows that the the Euler characteristic number is
$\chi(\mathcal{M})=2$, hence the apparent DM cannot alter the
topology of the space-time manifold, but can change the geometry
of the Universe in the sense that the curvature space scalar is
modified as
\begin{eqnarray}
k_{\star} =k (1-\zeta).
\end{eqnarray}
In particular, we can specialize the discussion for three cases:
For the close universe (with $k=+1$) we have $k_{\star}=(1-\zeta)>0$; for the
open universe (with $k=-1$) we have $k_{\star}=(\zeta-1)<0$; and finally, for the flat
universe, we get $k_{\star}=k(1-\zeta)=0$. This all shows that the
geometry is modified for the open and the closed universe,
provided $\zeta \neq 0$. For the perfectly flat universe with
$k=0$, we get no effect of $\zeta$. On the other hand, the effect
of $\Gamma \rho_B$ becomes important only at small scales and can
result with a bouncing universe as was shown in
\cite{Jusufi:2022mir}. In this emergent picture, the curvature of FLRW might not be a gauge-invariant quantity. The argument is the following: If the curvature is close to zero $k \approx 0$, (or close/open universe) in principle, we should expect a shift in curvature.  Since dark matter is an apparent effect, it is only present if there is baryonic matter, this for example, can be linked to the early phase when universe changed from a radiation-dominated era to a matter-dominated era. In addition, is is interesting to explore how the radiation matter changes the apparent dark matter. It might be that, apparent dark matter, is linked only to baryonic matter and not to radiation at all. We want to point out here that, it is interesting to explore a possible link with the "curvature tension" found in Planck data (see, \cite{DiValentino:2019qzk,Handley:2019tkm,Dhawan:2021mel,Vagnozzi:2020dfn}) and the effect of curvature shift found in our paper.

\section{Obtaining the $\Lambda$CDM model from Emergent gravity}
Before we obtain the $\Lambda$CDM model, let us  here assume a constant expression for the baryonic mass density which consists of normal matter, i.e,  $\rho_B=\rho_M=\rm const$ with $\bar{\beta}_B=0$, and $\rho_D=\rm const$ with $\bar{\beta}_D=0$, respectively. From Eq. (60) we get
\begin{eqnarray}
    \rho_D=\frac{2}{\sqrt{3}}\,\sqrt{\rho_M\,\rho_{\rm crit}},
\end{eqnarray}
where it has been defined
\begin{eqnarray}
    \rho_{\rm crit}=\frac{3 a_0}{8 \pi G R}=\frac{3 H_0^2}{8 \pi G },
\end{eqnarray}
where $R$ is assumed to be the Hubble radius. We can also write (see also \cite{Verlinde:2016toy})
\begin{eqnarray}
    \Omega_D=\frac{2}{\sqrt{3}}\,\sqrt{\Omega_M^{\Lambda CDM}}.
\end{eqnarray}
If we use data from \cite{Planck:2018vyg} with $\Omega_M^{\Lambda CDM} \simeq 0.05$ we get $\Omega_D \simeq 0.26$, which is quite remarkably. Now we can assume a generalization of the Eq. (83) by assuming that the densities depend on $a$ and evolve with time. In particular, in Eq. (83) we will assume that the critical density as a free parameter and then we define
\begin{eqnarray}
   \rho_D (a)=\frac{2}{\sqrt{3}}\sqrt{\alpha\,\rho_M^{\Lambda CDM}\,\rho_{\rm crit}} \,\,a^{-3}.
\end{eqnarray}
Note also that the apparent dark matter is a result of matter $\rho_M$. For the Friedman's equationwe have
\begin{eqnarray}\notag
\left(\frac{\dot{a}}{a}\right)^2+\frac{k}{a^2}&=&\frac{8\pi G}{3}\Big(\rho_R^{\Lambda CDM}(a)+\rho_M^{\Lambda CDM}(a)+\rho_D(a)\\
&+&\rho_\Lambda^{\Lambda CDM}\Big) \times \left[1-\sum_i\Gamma(\omega_i) \rho_i(a) \right].
\end{eqnarray}
here we not that we expect the Newton's constant to be in general $G \to G\,(1+\zeta)$, however the term can be absorbed in density parameters. This equation can be also written as
\begin{eqnarray}
    \Omega_D(z)=\frac{2}{\sqrt{3}} \sqrt{\alpha\, \Omega_M^{\Lambda CDM} } (1+z)^{3}.
\end{eqnarray}
This equation is one of the most important results in this paper. Firstly, there is a great level of similarity between the last expression and the expression for apparent dark matter obtained in Yukawa cosmology (see, Jusufi et al. \cite{Jusufi:2023xoa}). Secondly, Verlinde's relation (85) is obtained as a special case when $\alpha=1$. We can further use
\begin{eqnarray}
    \frac{a_0}{a(t)}=1+z,
\end{eqnarray}
with $a_0=1$. Further we can neglect the quantum gravity effect for the late universe and set $\Gamma \sim l_0^2 \sim 0$ (as they are expected to be small effects in large scales), in addition we are going to consider the Friedman's equation for the flat universe $(k=0)$, hence we can write
\begin{equation}
    \frac{H^2(z)}{H_0^2}=\Omega_R^{\Lambda CDM} a^{-4}+\Omega_M^{\Lambda CDM} a^{-3}+\Omega_D(a)+\Omega_\Lambda^{\Lambda CDM},
\end{equation}
or in terms of redshift $z$, we get
\begin{equation}\notag
  E^2(z)=\Omega_R^{\Lambda CDM}(1+z)^{4}+\Omega_M^{\Lambda CDM} (1+z)^{3}+\Omega_D(z)+\Omega_\Lambda^{\Lambda CDM},
\end{equation}
where it has been defined
\begin{eqnarray}
    E(z)=\frac{H(z)}{H_0}.
\end{eqnarray}
Finally, we can write
\begin{eqnarray}\notag
  E^2(z)&=&\Omega_R^{\Lambda CDM}(1+z)^{4}+\left(\Omega_M^{\Lambda CDM}+\frac{2\,\sqrt{\alpha\, \Omega_M^{\Lambda CDM} }}{\sqrt{3}}\right)\\
&\times& (1+z)^{3}+\Omega_\Lambda^{\Lambda CDM}.
\end{eqnarray}
In \cite{Jusufi:2023xoa}, it was shown that apparent dark matter, baryonic matter as well as cosmological constant are related via the simple equation \cite{Jusufi:2023xoa}
\begin{equation}
    \Omega_D=\sqrt{2\, \Omega_B^{\Lambda CDM} \Omega_{\Lambda}^{\Lambda CDM}} (1+z)^3.
\end{equation}
If we combine now this equation with Eq. (88) we obtain
\begin{eqnarray}
    \Omega_{\Lambda}^{\Lambda CDM}=\frac{2}{3}\alpha.
\end{eqnarray}
We can constrain the parameter $\alpha$ using observational data. On physical grounds, the parameter $\alpha$ describes the gravitational coupling between the baryonic matter and the quantum vacuum due to the vacuum entanglement. One can use the last equation and $\Omega_{\Lambda}^{\Lambda CDM}=\rho_{\Lambda}/\rho_{\rm crit}$, to show how explicitly the cosmological constant depends on alpha 
\begin{eqnarray}
    \Lambda=\frac{2 \alpha\,H_0^2}{c^2}.
\end{eqnarray} 
Having in mind that there exists a relation between the Hubble constant and the  Hubble scale via $H_0=c/L$ \cite{Verlinde:2016toy}, we can obtain 
\begin{eqnarray}
    \Lambda=\frac{2 \alpha}{L^2},
\end{eqnarray} 
and, similarly, we can obtain a holographic dark energy expression for the cosmological constant 
\begin{eqnarray}
    \rho_{\Lambda}=\frac{\alpha \,c^2}{4 \pi G L^2}.
\end{eqnarray} 
Here is also an interesting case of a very early universe with a constant density where the quantum effects play an important role. To see this and observe an interesting result let we set $\dot{a}=0$ (or equivalently $H=0$) in the Friedmann equation. Then, by solving the Friedmann equation we get the
equation for the critical density
$\rho_{\rm {c}}=(1 \pm \sqrt{2 G \pi
(2 G \pi a^2 -3 \Gamma  k}/2 G
\pi a )/2 \Gamma.$
This equation imposes a constraint $2 G \pi a^2 -3 \Gamma k  \geq 0,$
and consequently leading to the minimal scale factor $a=a_{\rm min}$, with  $a_{\rm min}=l_0\,\sqrt{2k}$ \cite{Jusufi:2022mir}. But since $l_0$ is of
Planck length order, i.e., $l_0 \sim l_{Pl}$, it follows that these effect become
important only in short distances. In such a case, for the critical density that
corresponds to the minimal scale one gets  $\rho_{\rm c} \sim
1/l_{Pl}^{2}$. This also shows that we end up with a bound for the maximum density in
nature. The cosmological implications of such a universe using the dynamical systems analysis depending on whether $-1 < \omega \leq -1/3$ and $\omega \geq -1/3$, and the equilibrium points of each system has been recently studied and discussed in great details (see, \cite{Millano:2023ahb}). If we identify $H_0$ with the value reported by the Planck collaboration, $H_0=67.40 \pm 0.50$ km s$^{-1}$ Mpc$^{-1}$
\cite{Planck:2018vyg}, and further take:
\begin{equation}
    \Omega_M^{\Lambda CDM}  \simeq 0.05,\,\, \Omega_R^{\Lambda CDM}  \simeq 9\,\,\times\,\,10^{-5},\,\,\,\Omega_\Lambda^{\Lambda CDM}  \simeq 0.69.
\end{equation}
Since our model here is effectively $\Lambda$CDM model with $\alpha \simeq 1.03$, to to compute the value reported by the Hubble Space Telescope (HST), $H=74.03 \pm 1.42$
km s$^{-1}$ Mpc$^{-1}$ \cite{Riess:2019cxk}, we can use the so-called look-back time quantity at a given redshift (see for details \cite{Capozziello:2023ewq}). In particular, one has to infer the value of the Hubble parameter using
\begin{equation}
   H=\frac{1+z}{T_0 z}\int_0^z \frac{dz'}{(1+z') E(z')}.
\end{equation} 
Using $T_0 \simeq 13.79$ Gyr and a fixed redshift $z$, we can obtain the value for the Hubble constant. This means that when $\alpha \simeq 1.03$, our results are same as in \cite{Capozziello:2023ewq}.

\section{Addressing the Hubble tension problem with evolving densities?}
In this final example, we point out that one can assume a different generalization of the Eq. (83) by assuming a different form for evolving densities. In particular, following the arguments in \cite{Verlinde:2016toy}, we can write
\begin{eqnarray}
    \rho_D(a)=\frac{2}{\sqrt{3}}\sqrt{\rho_B(a)\,\frac{3}{8 \pi G} \left(\frac{\dot{a}}{a}\right)^2}
\end{eqnarray}
Note also that $\rho_B=\rho_M+\rho_R$ includes the matter and also the radiation part, respectively. This equation can be also written as
\begin{eqnarray}
    \Omega_D(z)=\frac{2}{\sqrt{3}} \frac{H(z)}{H_0}\sqrt{\Omega_R (1+z)^{4}+\Omega_M (1+z)^{3}},
\end{eqnarray}
In terms of redshift $z$, and using the expression for $\Gamma$, we can write as follows
\begin{eqnarray}\notag
  \frac{H^2(z)}{H_0^2}&=&\left(\Omega_R(1+z)^{4}+\Omega_M (1+z)^{3}+\Omega_D(z)+\Omega_\Lambda\right)
  \\& \times & \left[1-\sum_i\Gamma(\omega_i) \rho_i(a) \right].
\end{eqnarray}
 We can neglect the quantum gravity effect and set $\Gamma \sim l_0^2 \sim 0$ (as they are expected to be small effects in large scales), we can finally write the last equation in terms of baryonic matter
 \begin{eqnarray}\notag
E^2(z)&=&\frac{2}{\sqrt{3}} E(z)\sqrt{\Omega_R (1+z)^{4}+\Omega_M (1+z)^{3}}\\
&+& \Omega_R(1+z)^{4}+\Omega_M (1+z)^{3}+\Omega_\Lambda.
\end{eqnarray}
This model is different and more general compared to the
$\Lambda$CDM model and it is interesting to see if it has any
relevance to the Hubble tension problem \cite{Giani}. More studies
on the Hubble tension can be carried out in
\cite{Valentino1,Valentino2,Valentino3,Pettorino1,Sunny1,Sunny2,Sunny3,Sunny4,Kami,Peri,Akar}.
We plan to study the phenomenological aspect of this model in a
separate paper.
%%%%%%%%%%%%%%%%%%%%%%%%%%%%%%%%%%%%%%%%%%%%%%%%%%%%%%%%%%%%%%%%%%%%%%%%%%%%%%%%%%%%%%%%%%%%%%%
\section{Conclusions}
In this paper, we have reported several new aspects of Verlinde's
Emergent Gravity (VEG) theory. First, we improved the formula of
the apparent DM in VEG theory by including the zero-point length
which encodes stringy effects. Importantly, we pointed out a close
correspondence between Verlinde's theory and non-local gravity.
This is inspired by the fact that the apparent DM is described by
entanglement entropy in VEG theory, however entanglement is a
non-local behavior of quantum mechanics, therefore there is no
surprise to say Verlinde's theory is a non-local gravity theory,
too. Next, we explored three different models for the apparent dark matter densities.
In the first model we studied a power law $\rho_D \sim R^{-2}$, ans we showed that the apparent DM modifies the classical
Friedmann equations in general relativity, in particular it shifts
the Newton constant of gravity in an interesting way coincides
with the MOG theory. Such a modification leads to important
results which play crucial role at small distances, for example to
cure the Big Bang singularities due to the minimal length and
large scale structures due to the apparent DM. Remarkably, we
showed that by rescaling the radial coordinate, the curvature
space constant, and the scale factor, then the classical
relativistic form of Friedmann equations and the FRW universe
metric remains invariant in terms of these new coordinates
provided the energy-momentum tensor has only baryonic matter.
Thus, since the curvature space constant is modified, it implies
that apparent DM can change the geometry of the universe since it shift the curvature space constant $k_{\star}=k (1-\zeta)$.  In our second model, we obtained the  $\Lambda$CDM model from emergent gravity. New relation for the density parameter of the apparent dark matter is obtained. This equation depends on a free parameter $\alpha$ and the Verlinde's relation is obtained as a special case when $\alpha=1$. In the final part, we study yet another model with evolving densities. By identifying $H_0$ with the value reported by the Planck
collaboration, $H_0=67.40 \pm 0.50$ km s$^{-1}$ Mpc$^{-1}$
\cite{Planck:2018vyg}, in the $\Lambda$CDM model one can use the look-back time quantity at a given redshift and given value of $\alpha$ parameter to address the problem of Hubble tension in the context of Verlinde's emergent gravity.  In the last section we obtained a more general model compared to  $\Lambda$CDM and we plan to study it's implications in a further work.
%%%%%%%%%%%%%%%%%%%%%%%%%%%%%%%%%%%%%%%%%%%%%%%%%%%%%%%%%%%%%%%%%%%%%%%%%%%%%%%%%%%%%%%%%%%%%%%%

\section{Acknowledgement} We are grateful to the Editor and the Referee for constructive comments which allowed us to  improve  significantly the paper.
SC acknowledges the support of {\it Istituto Nazionale di Fisica Nucleare} (INFN) ({\it iniziativa specifica}   QGSKY). 
This paper is based upon work from the COST Action CA21136, \textit{Addressing observational tensions in cosmology with systematics and fundamental physics} (CosmoVerse) supported by COST (European Cooperation in Science and Technology).

\end{document}